\documentclass[twocolumn,showpacs,preprintnumbers,amsmath,amssymb]{revtex4}
\usepackage{graphicx}% Include figure files
\usepackage{dcolumn}% Align table columns on decimal point
\usepackage{bm}% bold math
\usepackage{textcomp}
\usepackage{subfigure}

\newcommand\beq{\begin{equation}}
\newcommand\eeq{\end{equation}}
\newcommand\bseq{\begin{subequations}}
\newcommand\eseq{\end{subequations}}
\newcommand\bfig{\begin{figure}}
\newcommand\efig{\end{figure}}

\begin{document}

% the following line is for submission, including submission to the arXiv!!
\hspace{5.2in}% \mbox{Fermilab-Pub-04/xxx-E}

\title{Properties of discrete breathers in graphane from {\em ab initio} simulations}

%\affiliation{Research Institute of Physics, Southern Federal
%University, 194 Stachki Av, Rostov-on-Don 344090, Russia}
%
%\affiliation{Institute for Metals Superplasticity Problems of RAS,
%39 Khalturin St, Ufa 450001, Russia}
%
%\affiliation{National Research Tomsk State University, 36 Lenin
%Prospekt, Tomsk 634050, Russia}

%\author{G.M.~Chechin} \affiliation{Research Institute of Physics, Southern Federal University,
%194 Stachki Av, Rostov-on-Don 344090, Russia}
%\author{S.V.~Dmitriev} \affiliation{Institute for Metals Superplasticity
%Problems of RAS, 39 Khalturin St, Ufa 450001, Russia}
%\author{I.P.~Lobzenko} \affiliation{Research Institute of Physics, Southern Federal University,
%194 Stachki Av, Rostov-on-Don 344090, Russia}
%\author{D.S.~Ryabov} \affiliation{Research Institute of Physics, Southern Federal University,
%194 Stachki Av, Rostov-on-Don 344090, Russia}

\author{G.M.~Chechin$^1$, S.V.~Dmitriev$^{2,3}$, I.P.~Lobzenko$^1$, D.S.~Ryabov$^1$}

\affiliation{$^1$ Research Institute of Physics, Southern Federal
University, 194 Stachki Av, Rostov-on-Don 344090, Russia}

\affiliation{$^2$ Institute for Metals Superplasticity Problems of
RAS, 39 Khalturin St, Ufa 450001, Russia}

\affiliation{$^3$ National Research Tomsk State University, 36
Lenin Prospekt, Tomsk 634050, Russia}

\date{\today}

\begin{abstract}
A density functional theory (DFT) study of the discrete breathers
(DBs) in graphane (fully hydrogenated graphene) was performed. To
the best of our knowledge, this is the first demonstration of the
existence of DBs in a crystalline body from the first-principle
simulations. It is found that the DB is a robust, highly localized
vibrational mode with one hydrogen atom oscillating with a large
amplitude along the direction normal to the graphane plane with
all neighboring atoms having much smaller vibration amplitudes. DB
frequency decreases with increase in its amplitude, and it can
take any value within the phonon gap and can even enter the
low-frequency phonon band. The concept of DB is then used to
propose an explanation to the recent experimental results on the
nontrivial kinetics of graphane dehydrogenation at elevated
temperatures.
\end{abstract}

\pacs{63.20.Pw, 63.20.Ry, 63.20.dk, 63.22.Rc, 88.30.R-} \maketitle

\section{Introduction}
\label{Introduction}

Discrete breathers (DBs), also termed as intrinsic localized
modes, are spatially localized, large-amplitude vibrational modes
in defect-free nonlinear lattices. They have been identified as
exact solutions to a number of model nonlinear systems possessing
translational symmetry \cite{ST}. DBs were successfully observed
experimentally in various physical systems such as two-dimensional
array of optical waveguides \cite{Optics}; Bose-Einstein
condensate \cite{BoseEinstein}; one-dimensional micromechanical
array of coupled cantilevers \cite{Cantelivers}; two-dimensional
nonlinear electrical lattices \cite{English}; underdamped
Josephson-junction array \cite{Josephson}; quasi-one-dimensional
biaxial antiferromagnet \cite{Schwartz} and others.

Diversity of physical systems supporting DBs suggests that they
are very common in nonlinear lattices. Crystals are natural
nonlinear lattices and many studies, both experimental
\cite{Kalosakas,Kivshar,Uranium,Manley} and numerical
\cite{NaI,SiGe,NiNb,SavinKivshar,Shimada,WeGraphene,WeGraphane},
have been done to prove the existence of DBs and to use them for
explanation of various physical effects in crystals
\cite{Uranium,Bishop,Dubinko,Velarde,PRX,Juan}. Let us mention the
detection of DBs from the resonant Raman scattering measurements
in a complex compound termed as PtCl \cite{Kalosakas}; from
inelastic x-ray and neutron scattering data in $\alpha$-uranium
\cite{Uranium}; from inelastic neutron scattering spectra in NaI
\cite{Manley}. It should be noted that experimental observation of
DBs in crystals is a challenge because their contribution to the
vibrational density of states is masked by the contribution from
thermal lattice vibrations \cite{Critics}. In these circumstances
the importance of numerical studies cannot be overestimated.
Molecular dynamics based on empirical interatomic potentials was
used to identify DBs (or, more precisely, quasi-breathers
\cite{Chechin}) in NaI \cite{NaI}, in Si and Ge \cite{SiGe}, in Ni
and Nb \cite{NiNb}, in C$_{60}$ fullerite nanocrystals
\cite{SavinKivshar}; in carbon nanotubes \cite{Shimada}, graphene
\cite{WeGraphene,Khadeeva} and graphane \cite{WeGraphane}. In the
work \cite{GroupTheory} a group-theoretical approach has been
developed to simplify the analysis of existence and stability of
DBs of different symmetry types in nonlinear lattices. The
approach can also be applied to the analysis of DBs in
two-dimensional structures such as graphene and graphane.

Molecular dynamics studies rely on the quality of interatomic
potentials, which is always a question. For instance, the authors
of the work \cite{SiGe} report that they have tried different
interatomic potentials to model DBs in Si and have succeeded only
with the Tersoff potential. The reason is that the interatomic
potentials are often fitted to the elastic moduli and phonon
spectra of crystals (calculated from linearized equations of
motion) as well as to some experimentally measurable energies,
such as the sublimation energy, vacancy energy, etc. (for which
not the exact profile of the potential functions but their
integral characteristics are important since the change in
potential energy is path independent). On the other hand, DB,
being an essentially nonlinear vibrational mode, is sensitive to
the exact shape of the potentials. In this study it will be
demonstrated that the molecular dynamics simulation of DBs in
graphane \cite{WeGraphane}, performed using the LAMMPS package
\cite{50} with the AIREBO potential \cite{51}, gives an adequate
estimation of DB frequency only for small amplitudes and shows a
dramatic error for large amplitudes. At the same time, the AIREBO
potential has been successfully used in a countless number of
studies of various properties of hydrocarbon systems, meaning that
indeed DBs provide a very severe test of the interatomic
potentials.

Breather oscillations induce polarization of the outer electron
shells of atoms, which is very difficult to fully capture in frame
of the model considering interaction between mass points. There
exist several works where polarization of electron shells induced
by breather oscillations is partly taken into account. As an
example, we refer to the works \cite{Bishop} where a simplified
model was used to discuss the effect of polarization induced by
DBs in the perovskite structure.

The above discussion suggests the importance of {\em ab initio}
simulations of DBs in crystals. So far, to the best of our
knowledge, no such studies have been undertaken, possibly, because
the application of DFT theory to dynamical problems is
computationally costly. In this sense, graphane \cite{graphane,12}
is a very good choice for the study because the highly localized
DBs in a two-dimensional crystal can be analyzed using a
computational cell with a small number of atoms. Furthermore,
graphane is a new material promising for many applications
\cite{graphaneApplications}. Particularly, graphane holds a
potential for hydrogen storage due to its lightweight structure
and high performance \cite{12,graphaneApplications,35,37}. It was
shown that graphene can easily absorb hydrogen at low temperatures
and desorb at high temperatures \cite{12,19,25}.

Dehydrogenation kinetics during annealing of graphane turns out to
be not as simple as expected. It was found that there are two
types of dehydrogenation mechanisms with different dehydrogenation
barriers \cite{25}. It is challenging to find a theoretical
explanation of this effect.

Our motivation in this study is to legitimize the existence of
discrete breathers in crystalline solids by means of DFT
simulations using graphane as an example. We also aim to offer an
explanation of the dehydrogenation kinetics of graphane at
elevated temperatures.

The rest of the paper is organized as follows. In Sec.
\ref{Simulation} the simulation details a briefly presented, then
the results of DFT simulations are discussed in Sec.
\ref{Results}. The results are discussed in Sec. \ref{Discussion}
and summarized in Sec. \ref{Conclusions}.

\section{Simulation details}
\label{Simulation}

In the present work, DB in graphane is studied with the aid of
{\em ab initio} calculations. We use the ABINIT software package
\cite{Gonze} implementing the methods of the density functional
theory \cite{KS} (the corresponding algorithms can be found in
\cite{Payne}). The above package allows one to study dynamics of
molecules and crystals in the framework of the Born-Oppenheimer
approximation, which takes into account the significant
differences in the masses of the atom nuclei and electrons
\cite{BO}. The motion of heavy nuclei (ions) is described by the
classical equations, while that of light electrons is controlled
by the quantum mechanics equations. The forces acting on the
nuclei depend on the electronic subsystem state, which quickly
adapts to the nuclei current positions. Kohn-Sham equations
\cite{KS} are solved by ABINIT for each nuclei configuration. Our
calculations were performed based on the local density
approximation and Troullier-Martins pseudopotentials. Plane waves
are used as a basis set (with kinetic energy cutoff 40H) for the
decomposition of electron eigenstates.

\begin{figure}[t]
    \begin{center}
        \includegraphics[width=1.0\linewidth]{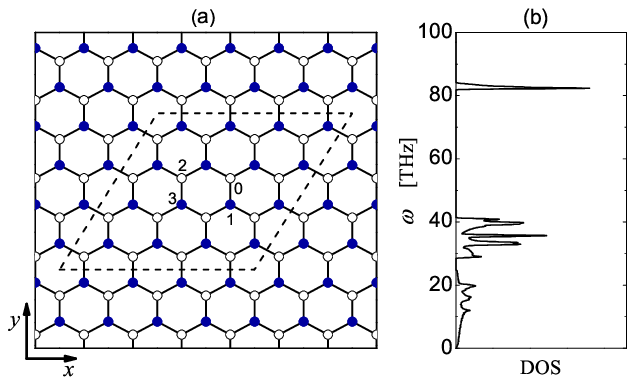}
        \includegraphics[width=1.0\linewidth]{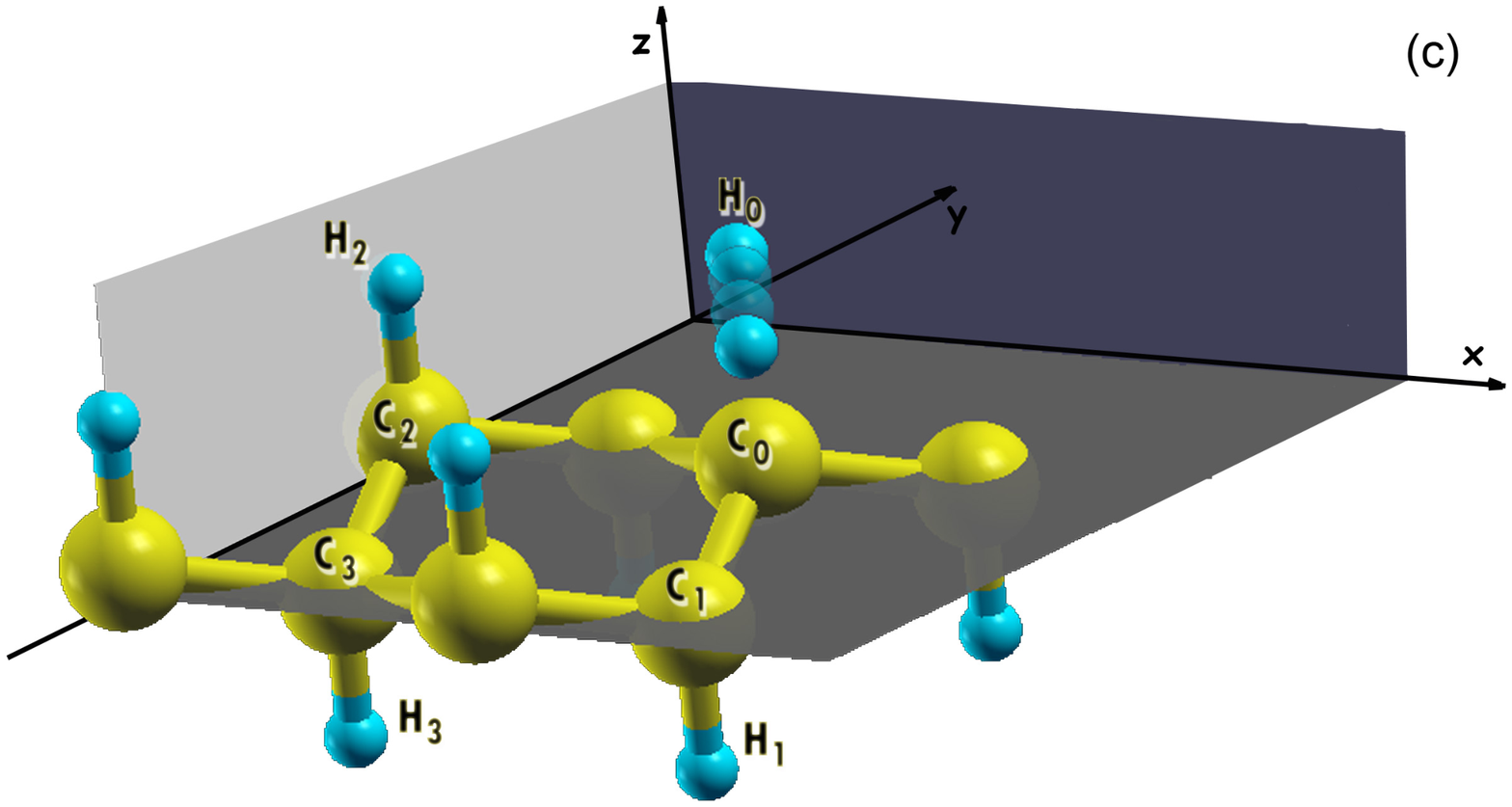}
        %\smallskip \medskip \bigskip \vspace[*]{length}
        \caption{(Color online) (a) Structure of graphane (CH). Open (filled) dots
        show carbon atoms with hydrogen atoms attached above
        (below) the sheet. The computational cell subjected to
        periodic boundary conditions is shown
        by the dashed line. To excite a DB, the H atom in the
        position 0, labeled as H$_0$, was displaced along $z$
        axis (normal to the graphane sheet) and released with zero
        initial velocity. All other atoms had zero initial
        displacements and velocities. (b) Phonon density of states (DOS) of
        graphane. (c) Schematic presentation of the DB in graphane.
        The H$_0$ atom vibrates with a large amplitude while all other
        atoms have much smaller vibration amplitudes.}
        \label{fig1}
    \end{center}
\end{figure}

The chair confirmation of graphane \cite{graphane} is considered
with H atoms attached at the opposite sides of the graphene sheet,
as shown in Fig. \ref{fig1}(a). Open (filled) dots show carbon
atoms with hydrogen atoms attached above (below) the sheet. The
computational cell, shown by the dashed line, includes 32 carbon
and 32 hydrogen atoms. Periodic boundary conditions are applied to
exclude the effect of free edges.

\section{Simulation results}
\label{Results}

Using the ABINIT package we have obtained the equilibrium
structure of graphane with the C-C bond length equal to 1.520~\AA,
and the C-H bond length equal to 1.117~\AA. Note that these
parameters coincide with those reported in \cite{graphane}. Due to
the interaction with hydrogen atoms, the carbon atoms split into
two sub-lattices with the off-plane displacements of
$\pm$0.228~\AA.

The calculated phonon density of states (DOS) for graphane is
shown in Fig. \ref{fig1}(b). The center of the narrow optical band
is at a frequency of about 83 THz, while the width of this band is
about 2 THz. The gap in the phonon spectrum extends from
$\omega_{\rm L}=41.7$ THz to $\omega_{\rm H}=81.6$ THz having the
width of 39.9 THz. These figures can be compared to the molecular
dynamics results of the work \cite{WeGraphane} where the phonon gap
edges were found to be 56.92 THz and 87.83 THz. Thus, the phonon
gap we found is about 20\% wider as compared to that given in Ref.
\cite{WeGraphane}. Overall, the linear vibration spectrum reported
in \cite{WeGraphane} is in a reasonable agreement with the {\it ab
initio} results of the present study.

Existence of a wide gap in the phonon spectrum of graphane opens
the possibility to excite a gap DB. This was achieved by applying
a displacement normal to the graphane plane (along $z$ axis) to
the H atom in the position 0, labeled as H$_0$ [see
Fig.~\ref{fig1}(a)]. All other atoms in the computational cell had
zero initial displacements and zero initial velocities. Varying
the initial displacement of H$_0$ atom, DBs with different
vibration amplitudes were excited. Figure~\ref{fig1}(c) gives a
schematic presentation of the DB in graphane. The H$_0$ atom
vibrates with a large amplitude while all other atoms have much
smaller vibration amplitudes.

\begin{figure}[t]
    \begin{center}
        \includegraphics[width=1.0\linewidth]{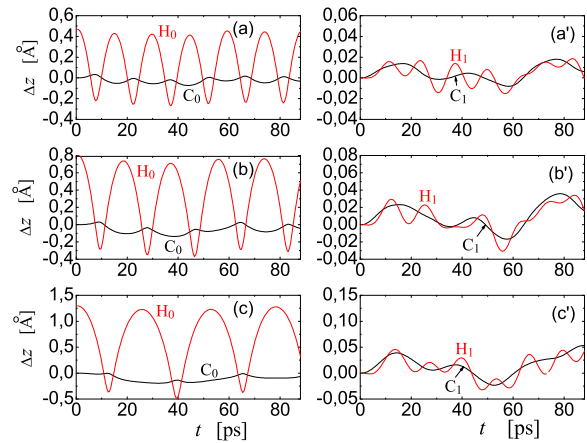}
        %\smallskip \medskip \bigskip \vspace[*]{length}
        \caption{(Color online) Displacements of atoms in the direction normal to the
        graphane plane as the functions of time. C$_0$, H$_0$ are
        the central atoms of DB and C$_1$, H$_1$ are their nearest neighbors
        [see Fig. \ref{fig1}(a)]. Black (red) lines show the displacements
        of C (H) atoms. Note different scale of ordinates
        used for the left and right panels. Amplitudes and frequencies of DBs are
        (a,a') $A=0.34$ \AA, $\omega_{\rm DB}=68.15$ THz;
        (b,b') $A=0.54$ \AA, $\omega_{\rm DB}=54.16$ THz;
        (c,c') $A=0.82$ \AA, $\omega_{\rm DB}=39.00$ THz.
        }
        \label{fig2}
    \end{center}
\end{figure}

In Fig. \ref{fig2} the $\Delta z$ displacements of the central
atoms of DB, H$_0$ and C$_0$ (left panels), as well as $\Delta z$
displacements of their nearest neighbors, H$_1$ and C$_1$ (right
panels), are shown as the functions of time. Black (red) lines
show the displacements of C (H) atoms. It can be seen that the
H$_0$ atom shows large-amplitude, quasi-periodic oscillations. The
C$_0$ atom vibrates with one order of magnitude smaller amplitude
because carbon is 12 times heavier than hydrogen. Excited DBs are
highly localized since the vibration amplitudes of the atoms H$_1$
and C$_1$ (and other atoms of the computational cell) are more
than one order of magnitude smaller than that of H$_0$ atom.

In order to quantify the DB amplitude and frequency the $\Delta
z(t)$ curve for H$_0$ atom is analyzed. The coordinates of the
successive minimum and maximum points on the curve, $\Delta
z_{\min}^{(n)}$, $t_{\min}^{(n)}$, $\Delta z_{\max}^{(n)}$,
$t_{\max}^{(n)}$, numbered by the index $n$ are determined. The
amplitude and the oscillation period for the $n$-th half
oscillation are defined as
\begin{eqnarray} \label{AW}
  A^{(n)}&=&(\Delta z_{\max}^{(n)}-
   \Delta z_{\min}^{(n)})/2, \nonumber \\
  \omega_{\rm DB}^{(n)}&=&(2 |t_{\max}^{(n)} -
  t_{\min}^{(n)}|)^{-1},
\end{eqnarray}
respectively. The quantities $A^{(n)}$ and $\omega_{\rm DB}^{(n)}$
averaged over a few tens of periods are taken as $A$ and
$\omega_{\rm DB}$.

\begin{figure}[t]
    \begin{center}
        \includegraphics[width=0.8\linewidth]{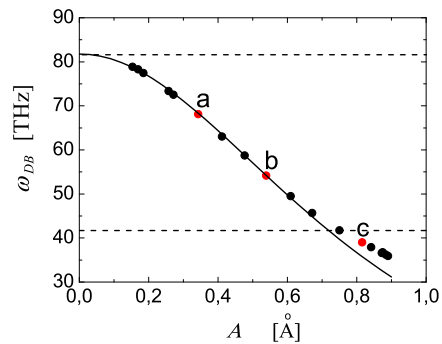}
        %\smallskip \medskip \bigskip \vspace[*]{length}
        \caption{(Color online) DB frequency as the function of amplitude
        from DFT simulation (scattered data). Solid line gives the analytical
        solution Eq. (\ref{AWMorse}) for a point mass in the Morse potential.
        Horizontal dashed lines show the edges of gap in the phonon density
        of states, $\omega_{\rm L}$ and $\omega_{\rm H}$. Dots labeled as
        a, b, and c (colored in red) correspond to the DBs presented in Fig.
        \ref{fig2} (a,a'), (b,b'), and (c,c'), respectively.}
        \label{fig3}
    \end{center}
\end{figure}

In Fig.~\ref{fig3} the DB frequency as the function of amplitude,
$\omega_{\rm DB}(A)$, is given. Edges of the phonon DOS gap are
shown by the horizontal dashed lines. Dots labeled as a, b, and c
(colored in red) correspond to the DBs presented in
Fig.~\ref{fig2} (a,a'), (b,b'), and (c,c'), respectively. It can
be seen that the $\omega_{\rm DB}(A)$ curve bifurcates from the
upper edge of the phonon gap and then decreases almost linearly
with increase in $A$, entering the lower phonon band. The decrease
in frequency with increase in amplitude reveals a soft-type
anharmonicity of the DBs in graphane in the entire range of DB
amplitudes.

\begin{figure}[t]
    \begin{center}
        \includegraphics[width=0.8\linewidth]{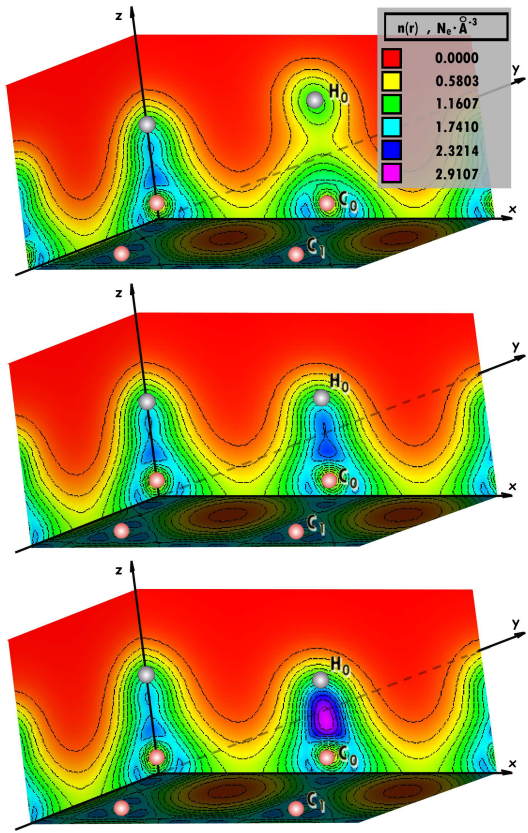}
        %\smallskip \medskip \bigskip \vspace[*]{length}
        \caption{(Color online) Electron density distribution in
        the vicinity of the DB for (a) maximal, (b) equilibrium
        and (c) minimal distance between the C$_0$ and H$_0$
        atoms constituting DB. The case shown in Fig.~\ref{fig2}~(a,a')
        (also marked by the red dot "a" in Fig.~\ref{fig3})
        is presented.}
        \label{fig4}
    \end{center}
\end{figure}

In Fig.~\ref{fig4} the electron density distribution is shown in
the vicinity of the DB for the case presented in
Fig.~\ref{fig2}~(a,a') and also marked by the red dot labeled as
"a" in Fig.~\ref{fig3}. Panels (a-c) of Fig.~\ref{fig4} correspond
to the maximal, equilibrium and minimal distance between the C$_0$
and H$_0$ atoms constituting DB. Three sections by the $(x,y)$,
$(x,z)$ and $(y,z)$ planes are shown. Within the presented
fragment the C$_0$, C$_1$, C$_2$, C$_3$, H$_0$ and H$_2$ atoms can
be seen (see Fig.~\ref{fig1}). It can be seen that due to the high
degree of localization of the considered vibrational mode, the
change in the C$_0$-H$_0$ bond length results in a noticeable
change in the electron density between these two atoms while it
has marginal effect on the electron density of other regions, for
instance, the C$_0$-C$_1$ and C$_2$-H$_2$ bond structures are
practically same in the panels (a) to (c).

\section{Discussion}
\label{Discussion}

Results of the present study can be compared to that in the
molecular dynamics study \cite{WeGraphane} where the $\omega_{\rm
DB}(A)$ curve was predicted to decrease for small $A$, to increase
for moderate $A$ and again to decrease at large $A$. Such a
non-monotonous $\omega_{\rm DB}(A)$ curve with DB frequency {\em
above} the phonon spectrum for large DB amplitudes is in a
striking difference with our {\em ab initio} result presented in
Fig.~\ref{fig3}.

According to the conventional definition, stationary DB is a
time-periodic dynamical object \cite{ST}. Excitation of a DB in
mathematical modeling requires the use of very refined initial
conditions \cite{Marine,FlachNumerical,soundvibr}. In the present
study a simple method was used for DB excitation and this is why
the DBs presented in Fig. \ref{fig2} are not exactly
time-periodic. The inaccuracy in the initial conditions results in
radiation of a part of energy given to the system at $t=0$ in the
form of small-amplitude lattice vibrations that disturb the
dynamics of DB. Spatially localized modes in nonlinear lattices
that do not show exact periodicity in time can be interpreted as
quasi-breathers \cite{Chechin}.

It is well-known that a stable DB should not have frequency within
the phonon spectrum of the crystal, otherwise it will excite the
extended normal modes and gradually loose its energy. Therefore,
the existence of the long-lived DBs with frequencies below
$\omega_{\rm L}$, presented in Fig.~\ref{fig2} (c,c'), may look
puzzling. The explanation is related to the fact that graphane has
rigidity with respect to in-plane deformation much higher than
bending rigidity. H$_0$ atom vibrating with a large amplitude
exerts a force normal to the graphane sheet producing primarily
bending deformation of the sheet. The bending phonon modes, due to
small bending rigidity, have frequencies much lower than
$\omega_{\rm L}$ and thus they are barely excited by the DB with
frequency just below $\omega_{\rm L}$. Similar arguments were used
to explain the existence of long-lived DBs and DB clusters in
strained graphene, having frequencies within the phonon band
\cite{WeGraphene,Khadeeva}.

Let us attempt to reproduced the $\omega_{\rm DB}(A)$ dependence
obtained for the DB in graphane (see Fig.~\ref{fig3}) using the
simplest one degree of freedom oscillator
\begin{eqnarray} \label{EqMotion}
  m\frac{d^2r}{dt^2}=-\frac{dU}{dr},
\end{eqnarray}
where $m=1.672\times 10^{-27}$ kg is the hydrogen atom mass,
$r(t)$ is the unknown coordinate of the atom as the function of
time and
\begin{eqnarray} \label{Morse}
  U(r)=D(1-e^{-ar})^2
\end{eqnarray}
is the Morse potential having minimum at $r=0$, the bond energy
$D$ and stiffness $a$. Equation of motion (\ref{EqMotion}) can be
solved exactly, giving the following relation between frequency
$\omega$ and amplitude $A=(r_{\max}-r_{\min})/2$
\begin{eqnarray} \label{AWMorse}
  \omega=\frac{a}{\pi}\sqrt{\frac{D}{2m}}{\rm sech}(aA).
\end{eqnarray}
The solid curve (\ref{AWMorse}) with the parameters $D=4.28$ eV
and $a=1.80$ \AA$^{-1}$ is plotted in Fig. \ref{fig3}. It can be
seen that the scattered data obtained with the use of DFT
simulations can be well fitted by the simple model in a wide range
of $A$. Note that the value of $D$ used for the fitting is in a
perfect agreement with the C-H bond dissociation energy of 4.25 eV
reported in \cite{Blanksby}.

We now turn to the discussion of recent experiments on thermally
activated dehydrogenation of graphane \cite{25}. Reportedly, there
are two activation energies of dehydrogenation with the transition
temperature at about 200$^{\rm o}$C. The smaller dehydrogenation
activation energy at temperatures below 200$^{\rm o}$C can be
understood by the metastable attachment of the energetic ions to
the graphene sheet during plasma hydrogenation \cite{19,25}.
However, the larger dehydrogenation activation energy at
temperatures above 200$^{\rm o}$C has not yet been well explained.
The nonlinear nature of the DBs implies that they do not appear at
relatively small temperatures and instead are excited at higher
temperatures. The fact that DBs can be externally excited at
temperatures near 0K indicates that they can also be spontaneously
excited at a finite-temperature thermal equilibrium
\cite{Lifetime,Ivanchenko}. DBs spontaneously excited at finite
temperatures may activate the dehydrogenation of hydrogenated
graphene.

The following arguments have been recently developed to modify the
reaction rate theory in solids taking into account DBs
\cite{Dubinko}. The classical expression for the reaction rate
$\dot{R}$ in thermal equilibrium reads
\begin{eqnarray} \label{Arrhenius}
  \dot{R}=R_0\exp\left(-\frac{E}{k_BT}\right),
\end{eqnarray}
where $R_0$ is the frequency factor having dimension of inverse
time, $E$ is the reaction activation energy, $k_B=8.617\times
10^{-5}$~eVK$^{-1}$ is the Boltzmann constant, and $T$ is
temperature. In their theory it is assumed that due to the
large-amplitude oscillations of atoms in the vicinity of DBs the
activation energy (but not the temperature) is a (quasi)periodic
function of time. For simplicity it is assumed that the height of
the potential barrier is harmonically modulated,
$E=E_0+e\cos(\Omega t)$, with the amplitude $e$ and frequency
$\Omega$. Even though the averaged over time potential barrier
height is unchanged, the modulation results in the following
correction of the escape rate \cite{Dubinko}
\begin{eqnarray} \label{ArrheniusCorrection}
  \dot{R}(e)=\dot{R}I_0\left(\frac{e}{k_BT}\right),
\end{eqnarray}
where $\dot{R}$ is the escape rate at $e=0$ and $I_0$ is the
zero-order modified Bessel function of the first kind. Note that
$I_0(0)=1$ and $I_0(x)$ is a rapidly growing function of $x$ for
$x>0$. For example, for $e=0.5$~eV at room temperature the
reaction rate amplification factor is of order of $10^7$, while
for $T=600$~K it is of order of $10^3$. This figures should be
multiplied by the DB concentration which increases with increase
in temperature according to the Arrhenius law
\cite{Lifetime,Ivanchenko}.

The above arguments suggest that DBs can considerably accelerate
thermal fluctuation-activeated chemical reactions such as
dehydrogenation of graphane.

\section{Conclusions}
\label{Conclusions}

In conclusion, the existence of DBs in a crystal was demonstrated
for the first time from {\em ab initio} calculations. Graphane was
chosen for this study because it supports highly localized gap
DBs, treatable by DFT simulations, and also because it is
promising for a number of applications.

DBs in graphane demonstrate the soft-type anharmonicity with
frequency monotonously decreasing with increasing amplitude. On
the other hand, molecular dynamics simulations based on the AIREBO
potential \cite{WeGraphane} give an adequate description of the
$\omega_{\rm DB}(A)$ curve only for small $A$ and fail for large
$A$. These results suggest that DBs provide a very severe test of
the interatomic potentials used in molecular dynamics simulations.

The concept of DBs together with the reaction-rate theory that
takes into account DBs were used to explain basic physics behind
the dehydrogenation kinetics of graphane at finite temperatures.

Our work opens new direction for further research, which is
analysis of properties of DBs in crystals by means of {\em ab
initio} simulations.

The research was partly supported by the Russian Science
Foundation (grant No14-13-00982) and by the Russian Foundation for
Basic Research, grants 12-02-31507-mol-a and
14-02-97029-p-Povolzhie-a. The present results have been obtained
through the use of the ABINIT code, a common project of the
Universit\'e Catholique de Louvain, Corning Incorporated, and
other contributors (URL http://www.abinit.org).

\end{document}